\documentclass[sn-mathphys-num]{sn-jnl}

\usepackage{amsmath}
\usepackage{empheq}
\usepackage{mathrsfs}
\usepackage{geometry}                
\usepackage{graphicx}
\usepackage{amssymb}
\usepackage{epstopdf}
\usepackage{graphicx}%
\usepackage{multirow}%
\usepackage{amsmath,amssymb,amsfonts}%
\usepackage{amsthm}%
\usepackage{mathrsfs}%
\usepackage[title]{appendix}%
\usepackage{xcolor}%
\usepackage{textcomp}%
\usepackage{manyfoot}%
\usepackage{booktabs}%
\usepackage{algorithm}%
\usepackage{algorithmicx}%
\usepackage{algpseudocode}%
\usepackage{listings}%
\usepackage{natbib}%
\usepackage{mathabx}
\usepackage{wasysym}

\newcommand{\G}{\mathcal{G}}

\newcommand{\M}{M_{\jupiter}}
\newcommand{\Mjup}{M_{\jupiter}}
\newcommand{\Rjup}{R_{\jupiter}}
\newcommand{\Ijup}{I_{\jupiter}}
\newcommand{\Rt}{R_{t}}
\newcommand{\Rc}{R_{c}}

\begin{document}

\title[]{\textbf{Determination of Jupiter's Primordial Physical State}} 

\author*[1]{\fnm{Konstantin} \sur{Batygin}}\email{kbatygin@caltech.edu}

\author[2,3]{\fnm{Fred C.} \sur{Adams}\email{fca@umich.edu}}

\affil*[1]{\orgdiv{Division of Geological and Planetary Sciences}, \orgname{California Institute of Technology}, \orgaddress{\street{1200 E. California Blvd.}, \city{Pasadena}, \postcode{91125}, \state{CA}, \country{USA}}}

\affil[2]{\orgdiv{Physics Department}, \orgname{University of Michigan}, \orgaddress{\street{450 Church St}, \city{Ann Arbor}, \postcode{48109}, \state{MI}, \country{USA}}}
\affil[3]{\orgdiv{Astronomy Department}, \orgname{University of Michigan}, \orgaddress{\street{450 Church St}, \city{Ann Arbor}, \postcode{48109}, \state{MI}, \country{USA}}}

\maketitle

\textbf{The formation and early evolution of Jupiter played a pivotal role in sculpting the large-scale architecture of the solar system, intertwining the narrative of Jovian early years with the broader story of the solar system's origins. The details and chronology of Jupiter’s formation, however, remain elusive, primarily due to the inherent uncertainties of accretionary models, highlighting the need for independent constraints. Here we show that by analyzing the dynamics of Jupiter's satellites concurrently with its angular momentum budget, we can infer Jupiter's radius and interior state at the time of proto-solar nebula’s dissipation. In particular, our calculations reveal that Jupiter was $2$ to $2.5$ times as large as it is today, 3.8 million years after the formation of the first solids in the solar system. Our model further indicates that young Jupiter possessed a magnetic field of approximately $B_{\jupiter}^{\dagger} \approx 21$ mT (a factor of $\sim50$ higher than its present-day value) and was accreting material through a circum-Jovian disk at a rate of $\dot{M} = 1.2-2.4$ Jupiter masses per million years. Our findings are fully consistent with the core-accretion theory of giant planet formation and provide an evolutionary snapshot that pins down properties of the Jovian system at the end of the protosolar nebula's lifetime.}

\section{Introduction}

A restricted view of the solar system -- wherein a multitude of bodies evolve subject to the sole influence of Jupiter and the Sun -- is a well-established and long-standing paradigm of celestial mechanics \citep{Poincare1892,1963RuMaS..18...85A, 1997CMaPh.186..413C}. Despite being grossly simplified, the analogy between this picture and reality has only deepened in time, as the theory of planet formation gradually illuminated Jupiter's decisive role in sculpting the solar system's large-scale architecture \citep{Walsh2011, BL15, 2017PNAS..114.6712K}. Accordingly, a full delineation of Jovian structure and origins is often viewed as a key milestone on the path toward unveiling the broader narrative of the solar system’s primordial evolution \citep{2022Icar..37814937H, 2023ASPC..534..947G}. Though considerable uncertainties regarding the interior persist (in part due to gravity data's inability to fully inform the nature of the compact core as well as uncertainties in the equation of state of hydrogen itself \citep{Miguel2022, Howard2023}) over the course of the past three decades, insights from the Galileo and Juno missions have brought Jupiter's complex, multi-layered interior into sharper focus. In particular, recent modeling efforts have revealed a region permeated by Helium rain, a dilute high-metallicity core of up to 25 Earth masses, along with a considerably less massive, deep-seated compact core \citep{MH24}. In contrast with these advances, a complete understanding of Jupiter's formation remains elusive.

Generally speaking, the physical characteristics of Jupiter align with the predictions of core-accretion (also known as the core-nucleated instability) model of giant planet formation \citep{1980PThPh..64..544M,1982P&SS...30..755S,1986Icar...67..391B}. Within this framework, the formation of giant planets follows a distinct series of stages. Initially, a high-metallicity core forms rapidly, giving way to a period of hydrostatic growth characterized by a slow agglomeration of a H/He atmosphere. This process continues until the gaseous envelope's mass matches that of the core. Once this threshold is crossed, a transient period of rapid gas accretion ensues, facilitating the accumulation of most of the planet's mass. Eventually, the planet separates from the surrounding nebula, embarking on a path of long-term thermal evolution that results in the Jupiter we observe today, approximately 4.5 billion years later \citep{Pollack96}.

Although the broad strokes of this picture have been established for decades, the intricacies of Jupiter's early evolutionary sequence remain imperfectly understood. In particular, Jupiter’s primordial entropy -- often framed as the ``hot versus cold start" problem -- alongside the exact timing of its formation phases remain elusive. For instance, within the the oft-cited model of \citet{Pollack96}, the transition to runaway accretion occurs approximately 7 million years after core formation. Subsequent calculations \citep{2005Icar..179..415H,2013A&A...558A.113M, 2021Icar..35514087D,2023A&A...675L...8H}, however, have proposed alternative chronologies, with recent model of \citet{Stevenson22}, suggesting that runaway growth concludes by the $t=3\,$Myr mark. Additionally, the role of shock cooling, double-diffusive convection, and other effects \citep{2018MNRAS.477.4817C, 2012A&A...540A..20L} introduce further complexity to a first-principles approach to this problem.

\begin{figure*}[t!]
\centering
\includegraphics[width=0.85\textwidth]{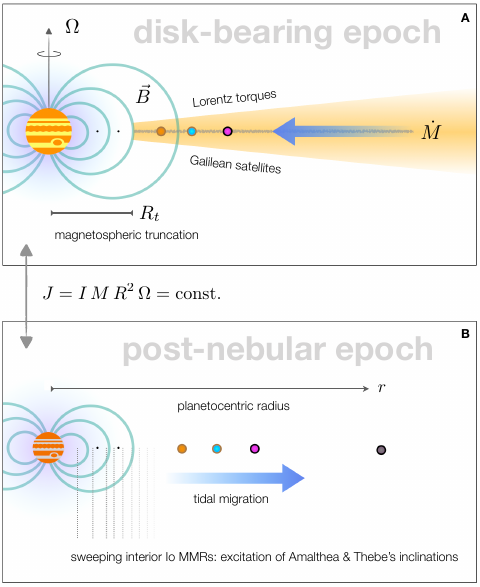}
\caption{A schematic overview of the key processes at play in our model. Panel (A) depicts an epoch of Jovian evolution close to the end of the lifetime of the proto-solar nebula (approximately 4 Myr after formation of the first solids within the solar system). The circum-Jovian disk is represented by the orange shaded area and is truncated by the magnetosphere (pale blue lines) at a planeto-centric radius $r=\Rt$. The primordial spin rate of Jupiter, $\Omega$, is regulated by Lorentz torques, and lags co-rotation at $\Rt$ by a small, well-defined amount. The coloured circles represent the Galilean satellites (orange: Io; cyan: Europa; magenta: Ganymede) and the small dots inside the magnetosphere denote the moons Amalthea and Thebe. The blue arrow indicates the accretionary flow of disk material, which occurs at a rate $\dot{M}$. Panel (B) portrays the Jovian system at a later epoch. Io’s outward tidal migration sweeps its interior mean-motion resonances across Amalthea and Thebe, inducing mild excitation of their orbital inclinations. The system’s post-nebular evolution is governed by the near-exact conservation of Jupiter's rotational angular momentum, $J=I\,M\,R^2\,\Omega$ where $I$ is the dimensionless moment of inertia, $M$ is Jupiter’s mass, and $R$ is its radius.}
\label{f1}
\end{figure*}

Identifying a diagnostic of Jovian physical state at a given moment in time, without resorting to traditional accretion models would provide a critical, independent constraint on its formation process. In this work, we propose leveraging the early dynamics of Jupiter's satellites, alongside magnetic regulation of its angular momentum budget, to infer Jupiter’s radius -- and thereby its interior state -- at the time of nebular dissipation. This approach -- summarized schematically in Figure \ref{f1} -- largely bypasses the limitations of existing models, and casts the properties of the Jovian system at its formative stages into an unprecedentedly sharp focus.

\section{Results: Jupiter's Primordial Radius}

Interior to the famed Galilean satellites, Jupiter is encircled a system of smaller moons: Thebe, Amalthea, Adrastea and Metis (in order of diminishing planeto-centric distance). With physical radii of only $20\,$km and $8\,$km, the latter pair -- Metis and Adrastea -- are considerably smaller than the former, and orbit the planet within the fluid Roche limit, separated by merely $\sim1000\,$km in their semi-major axes. These attributes suggest that Adrastea and Metis are ``ring-moons" that originated from the disintegration of a larger precursor, suggesting that their current orbits may not be primordial \citep{Burns}. In contrast, Amalthea and Thebe are a factor of $\sim10-1000$ more massive, and occupy orbits that lie further away from Jupiter. The most natural explanation for their existence is that they were resonantly transported inward by Io, during its phase of disk-driven in-spiral \citep{BM20}. Although the exact details of these moons' origins are unimportant for our purposes, this scenario is consistent with the detection of a $\sim3\mu\rm{m}$ absorption feature in Amalthea's surface spectrum \citep{Takato2004}, which implies that it was transported inward from a larger initial orbit within the circum-Jovian disk. The orbits of Amalthea and Thebe are thus almost certainly primordial, and in fact, hold a clear dynamical footprint of Io's post-nebular tidal migration.

The trajectories of Amalthea and Thebe lie close to the Jovian equatorial plane, but their inclinations are measurably non-zero, clocking in at $0.36 \deg$ and $1.09 \deg$, respectively. The work of \citet{Hamilton} has convincingly demonstrated that these specific values are consequences of divergent encounters of the satellites' orbits with second-order inclination resonances with Io. That is, as Io tidally migrated outward from its primordial semi-major axis, $a_{\rm{Io}}^{\dagger}$ (throughout this work we adopt a notation where the dagger$^\dagger$ superscript refers to primordial values, evaluated at the time of disk dissipation), the inner satellites crossed mean-motion commensurabilities, which delivered impulsive gravitational excitations of their orbits. By matching the magnitudes of the resulting inclination kicks to those expected from Io’s migration, \citet{Hamilton} showed that Io must have began its orbital evolution at a semi-major axis that exceeded the present-day Jovian radius by a factor of $\xi = a_{\rm{Io}}^{\dagger}/\Rjup = 4.02-4.98$. The physical origins of these limits are intelligible: had Io begun its migration interior to $4.02\Rjup$, Amalthea would have been swept by the 4:2 resonance, resulting in a higher-than-observed inclination. Conversely, an explanation for Thebe's inclination, at the very least, necessitates passage through the 6:4 commensurability, yielding the upper bound. We further note that while the full quoted range of $\xi$ is technically plausible, values closer to the lower limit -- $4.02 - 4.06\Rjup$ -- are more likely, given that Thebe's inclination is most naturally explained by invoking sequential encounters with 6:4, 5:3, as well as 4:2 resonances (Methods section \ref{app:dynamics}; Figure \ref{sfmr}).

Importantly, $a_{\rm{Io}}^{\dagger}$ does not correspond to Io’s \textit{formation} radius -- merely its orbital radius at the epoch of nebular dissipation. Modern models of satellite formation broadly agree that \textit{in-situ} accretion of the Galilean moons is highly unlikely, and that they must have formed at a much greater orbital radius, before migrating inward and stabilizing near the disk’s inner edge \citep{2002AJ....124.3404C,BM20,2020A&A...633A..93R}. The latter point is of considerable importance: the existence of a magnetospheric cavity in the circumplanetary disk generates a sharp transition in the surface density, which greatly enhances the corotation torque exerted onto the satellite by the gas \citep{Masset2006, 2011MNRAS.410..293P}, thereby stalling -- or even reversing -- disk-driven (type-I) migration. A suite of numerical simulations \citep{Masset2006,AtaieeKley21} have self-consistently shown that once trapped near the disk’s inner edge, the equilibrium orbital radius of the inner-most member of a resonant chain composed of three objects with a secondary-to-primary mass ratio similar to the Io-Europa-Ganymede triplet, exceeds the disk’s truncation radius, $\Rt$, by a factor of $\zeta = a_{\rm{Io}}^{\dagger}/\Rt \approx 1.13$ (see Methods section \ref{app:migration} for a discussion). Consequently, at the time of nebular dissipation, the circum-Jovian disk must have been truncated at a radius of $\Rt \approx 3.6-4.4\, \Rjup$. 

The same process that facilitates disk truncation (namely magnetohydrodynamic coupling between the planetary magnetic field and the circumplanetary nebula) also regulates the terminal angular momentum budget of the planet \citep{Bat18}. That is, as a consequence of magnetic induction and reconnection within the disk, the planetary spin tends towards an equilibrium state, given by $\Omega_{\Jupiter}^{\dagger} = \chi\,\sqrt{\G\,\Mjup/\Rt^3}$, where $\chi\approx(1/2)\sqrt{2+3/(\sqrt{2}\,\alpha)} = 0.88$ is a constant that is largely determined by the balance between accretionary spin-up and magnetic breaking (Methods section \ref{app:rotation}). Notably, this quantity is insensitive to both the field strength and accretion rate, and only weakly depends on a order-unity constant, $\alpha$, which is in turn dictated by field geometry.

\begin{figure*}[t!]
\centering
\includegraphics[width=0.85\textwidth]{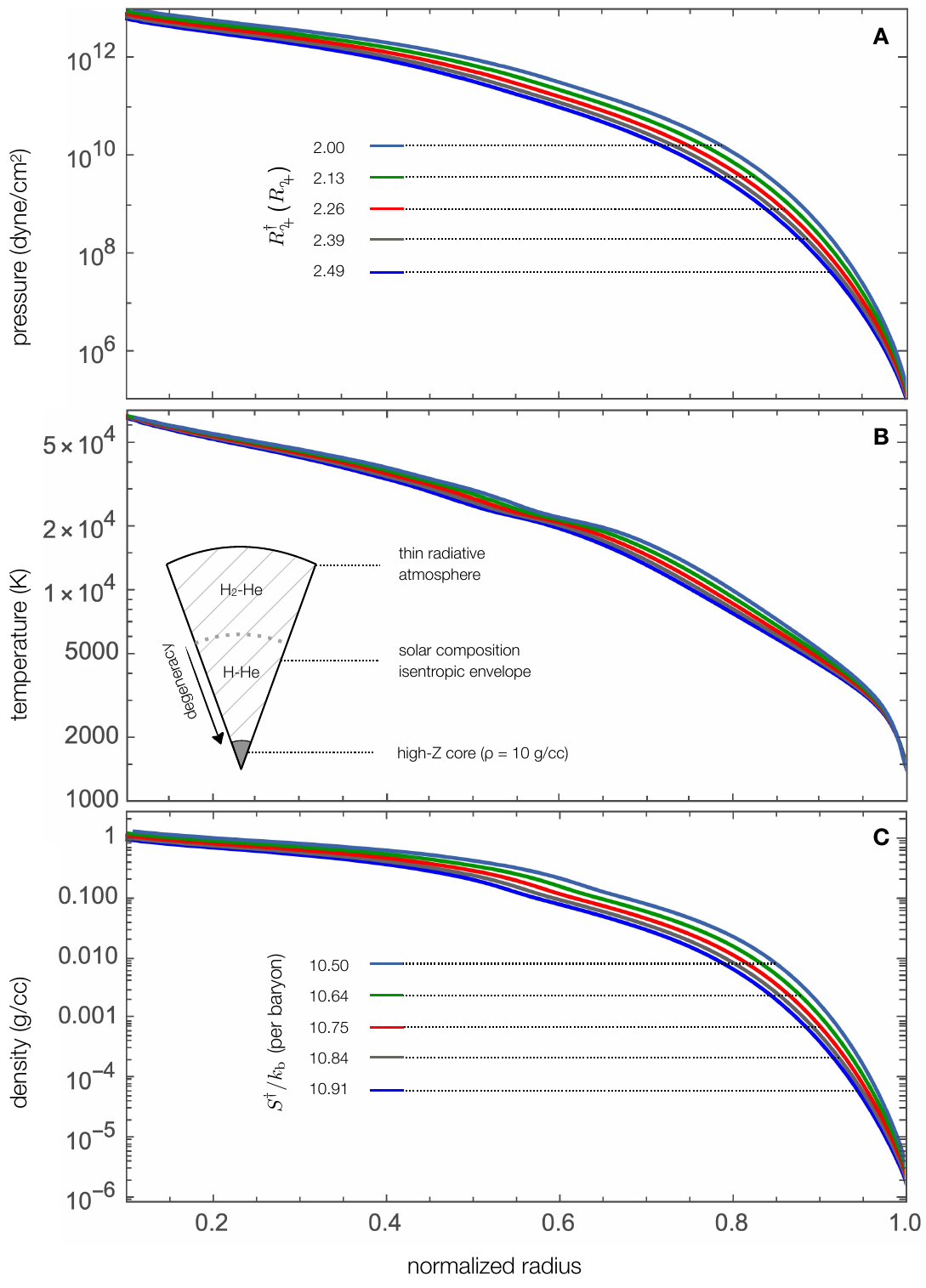}
\caption{Hydrostatic models of Jupiter's interior structure at the time of nebular dissipation. The three families of multi-colored curves show profiles of the pressure (panel A), temperature (panel B), and density (panel C) as functions of the distance away from the planet's center, normalized by the physical radius, $\Rjup^{\dagger}$. The calculations assume the existence of a high-metallicity $M_{\rm{c}} = 25 M_{\oplus}$ core and a solar composition envelope, consistent with the inferred total heavy element budget of Jupiter \citep{MH24}. The depicted models span the admissible range of Jupiter's primordial radius, and are color-coded accordingly, along with the corresponding interior entropy of the convective envelope.}
\label{f2}
\end{figure*}

This rotational state is maintained until the eventual evaporation of the circum-Jovian nebula. Following its dissipation, Jupiter's spin increases as a result of gravitational contraction -- a process that adheres to the conservation of rotational angular momentum, $J=I\,M\,R^2\,\Omega$. To this end, we note that tidal migration of the Galilean satellites extracts a negligible fraction of the planet's spin angular momentum, and can be safely neglected. The current value of $J$ is well-established \citep{Helled2011MOI,Wahl2017,Ni2018}, and the initial spin rate is constrained by the inferred truncation radius of the circum-Jovian disk. Furthermore, the primordial moment of inertia factor of Jupiter, $\Ijup^{\dagger}$ is readily calculated as a monotonic function of the planetary radius utilizing conventional structure models (Methods section \ref{app:interior}; Figure \ref{fmesa}). In this work, we adopt a simple ``layered" structure model -- wherein a heavy-element core is encompassed by a fully convective solar-composition envelope -- neglecting the notion that Jupiter likely formed with compositional gradients in the deep interior \citep{Lozovsky2017,Vazan2018,VallettaHelled2019,Helled2022}. As we show in Methods section \ref{app:interior}, however, the determination of $\Ijup^{\dagger}$ is not particularly sensitive to the detailed character of the assumed interior structure. Thus, by rearranging the law of angular momentum conservation, we obtain an equation for Jupiter's radius at the epoch of nebular dissipation:
\begin{align}
\Rjup^{\dagger} = \Bigg( \frac{\Ijup}{\Ijup^{\dagger}} \frac{\Omega_{\Jupiter}}{\chi\,\Omega_{\rm{br}}}  \sqrt{ \frac{\xi^3}{\zeta^3} } \Bigg)^{1/2}\,\Rjup,
\label{yes}
\end{align}
where $\Omega_{\rm{br}} = \sqrt{\G\,\Mjup/\Rjup^3}$ is the present-day Jovian breakup rotation rate. 

Given the values of $\chi$, $\xi$ and $\zeta$ quoted above, and self-consistently expressing $\Ijup^{\dagger}$ as a function of $\Rjup^{\dagger}$, we obtain the following range for Jupiter's primordial radius: $\Rjup^{\dagger} = 2.02-2.59\,\Rjup$. Within the range spanned by our estimate of $\Rjup^{\dagger}$, any value that exceeds the orbital radius of Amalthea is unlikely to be physically meaningful. Along similar lines, we note that because Thebe's inclination is probably a result of multiple resonance passages, the lower end of the derived radius range is favored anyway \citep{Burns}. Finally, though the values of the order-unity dimensionless constants in equation (\ref{yes}) can in principle be altered to some extent, the analytic nature of this estimate makes this straightforward, and given that the dependence is sub-linear (e.g., such as square-root and quarter-root for $\zeta$ and $\alpha$ respectively), these adjustments are unlikely to significantly affect the overall result.

With the planetary radius constrained, we infer the interior structure of Jupiter using the same hydrostatic models that we employed to parameterize $\Ijup^{\dagger}$ as a function of $\Rjup^{\dagger}$. The corresponding temperature, density, and pressure profiles are shown in Figure \ref{f2}. Intriguingly, these models imply that at the time of proto-solar nebular dissipation, the characteristic entropy of Jupiter's convective envelope is constrained to $S_{\jupiter}^{\dagger}\approx 10.6-11\,k_{\rm{b}}$ per baryon, corresponding to a ``warm start" initial condition \citep{2012ApJ...745..174S}. 

The derived entropy estimate is broadly consistent with previous works, although a wide range of parameters are possible. For example, Berardo \& Cumming \cite{berardo2017} find that young Jupiter-mass planets have radius $R \sim 2\,\Rjup$ and entropy of $S \approx 11 \,k_{\rm{b}}$ per baryon for a shock temperature of $2500\,$K, with the entropy deceasing to $S \sim 10 \,k_{\rm{b}}$ per baryon after 4 Myr of evolution. A distinct analysis \citep{marley2007} suggests that newly-formed Jupiter-type planets have somewhat smaller entropy $S\sim9-10 \, k_{\rm{b}}$ per baryon for both ``hot start" models and conventional accretion models, though the entropy estimates increase significantly with mass for the ``hot" case. A number of other studies \citep{marley2007,mordasini2013,mordasini2017,berardoetal2017,marleau2019} indicate entropies in an even more expanded range, spanning $S\sim\,10-14 \,k_{\rm{b}}$ per baryon. This wide range of possible properties is driven by the action of on ancillary processes, such as the behavior of the accretion shock during gas infall. Additionally, the density profile of the planet was almost certainly more complex than portrayed by our simple model, including a core that was partially mixed with the envelope, as well as regions where convection may have been suppressed \citep{VallettaHelled2019,Helled2022}. As a result, future models should further refine our estimation of Jupiter's primordial entropy by fusing the constraints derived herein with more sophisticated models of the interior structure.

\medskip

A final piece of this puzzle lies in pinning down the above calculation to a specific epoch, relative to a well defined marker in the solar system’s evolution. In this vein, \citet{2017Sci...355..623W} used Angrite magnetization data to demonstrate that the solar nebula dissipated $\sim3.8\,$ Myr after CAI (Calcium-Aluminum Inclusion) formation. Because the energy required to transport a hydrogen molecule from a few Jovian radii to the Hill radius is approximately equal to that required to remove it from the gravitational potential well of the sun at 5AU altogether (i.e., $\G \M/a_{\jupiter}\sim\G \Mjup/(10\,\Rjup)$), the photo-evaporative front responsible for blowing away the nebula in the Jovian orbital neighborhood is bound to have simultaneously removed the circumplanetary disk as well. Thus, we conclude that Jupiter was approximately $2-2.5$ times as large as it is today, 3.8 million years after the formation of the first solids in the solar system.

\section{Discussion: Jupiter's Primordial Magnetic Field and Mass-Accretion Rate}

\begin{figure*}[t]
\centering
\includegraphics[width=0.85\textwidth]{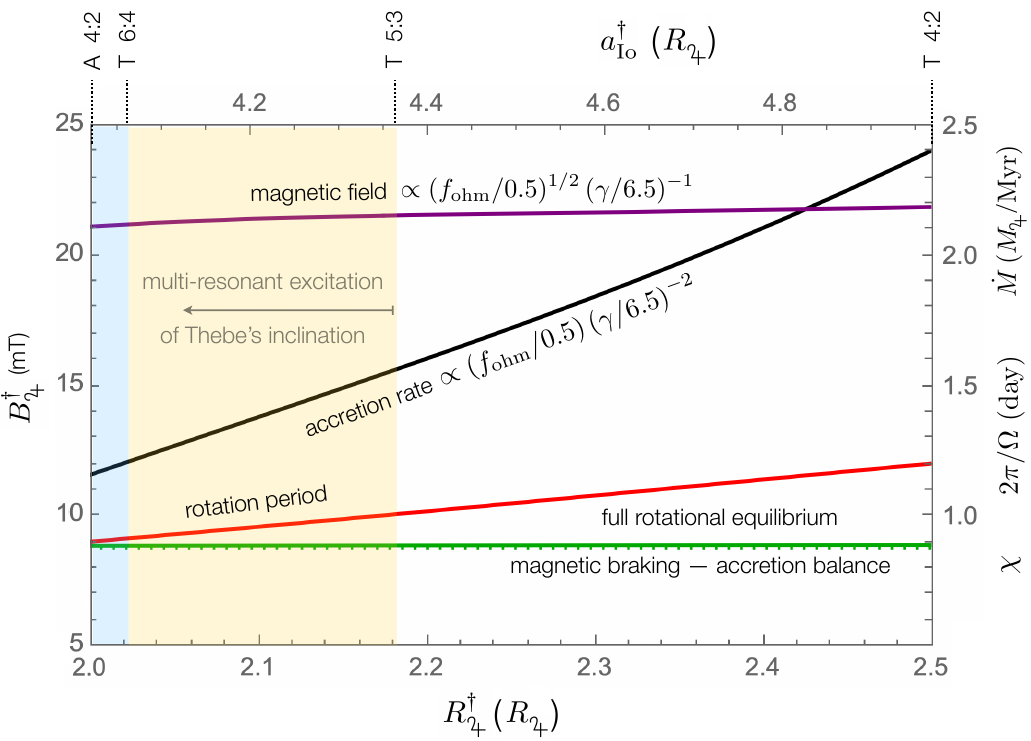}
\caption{Inferred primordial Jovian magnetic field ($B_{\jupiter}^{\dagger};$ purple curve, left axis), as well as the mass-accretion rate ($\dot{M}$; black curve, right axis) and the rotation period ($2\pi/\Omega;$ red curve, right axis), plotted as functions of Jupiter's primordial radius ($R_{\jupiter}^{\dagger}$). The ratio of Jovian rotation frequency to the orbital frequency at the truncation radius ($\chi$; green lines, right axis) is also shown, where the dotted curve is the analytic estimate (see Methods Section~\ref{app:rotation}) and the solid curve is obtained from a self-consistent rotational equilibrium solution. In the quoted scaling relations, $f_{\mathrm{ohm}}$ refers to the ratio of Ohmic to total energy dissipation in Jupiter’s interior, and $\gamma$ is the ratio of the mean internal field to the surface field. Although $\dot{M}$ and $B_{\jupiter}^{\dagger}$ are based on the most plausible values of $\gamma$ and $f_{\mathrm{ohm}}$, they represent effective lower bounds. On the top axis, $a_{\mathrm{Io}}^{\dagger}$ denotes the primordial orbital radius of Io, and the letters ``A'' and ``T'' at the top mark the radii at which resonances with Amalthea and Thebe occur, respectively. The shaded regions highlight the favored range of $R_{\jupiter}^{\dagger}$ in which Thebe’s orbital inclination is excited by sequential resonances (yellow for two resonances, cyan for three).}
\label{f3}
\end{figure*}

Our estimates of the truncation radius of the circum-Jovian disk, as well as the planetary radius itself, offer an additional avenue towards characterizing the rate of mass accretion facilitated by the disk. In its simplest form, the expression for the truncation radius of the circum-Jovian disk can be obtained by equating the magnetic pressure exerted by Jupiter's field to the accretionary ram pressure within the disk \citep{GhoshLamb78}. This conventional derivation has been validated by the recent observations \citep{2023ApJ...958..140C} and simulations of \citet{Zhu24}, which confirm its accuracy from self-consistent numerical grounds. Importantly, the resulting expression for $\Rt$ depends on the planetary radius, the magnetic field strength, as well as the disk's accretion rate.

For rapidly rotating, fully-convective astrophysical dynamos, strength of the generated surface field is governed by an equipartition-like relationship \citep{Christensen2009}: 
\begin{align}
B_{\jupiter}^{\dagger} = \frac{\sqrt{2\,\mu_0\,c\,f_{\rm{ohm}}}\,\big(\mathcal{F}\,q \big)^{1/3}\,\langle \rho \rangle^{1/6}}{\gamma} \approx 21.2\,\mathrm{mT} \, \bigg(\frac{f_{\rm{ohm}}}{0.5} \bigg)^{1/2}\bigg(\frac{\gamma}{6.5} \bigg)^{-1}\,\bigg( \frac{\Rjup^{\dagger}}{2\,\Rjup} \bigg)^{1/6}
\label{yesyes}
\end{align}
where $c=0.63$ is a constant of proportionality, $f_{\rm{ohm}}\approx 0.5$ is the ratio of Ohmic to total energy dissipation, and $\mathcal{F}\sim 1$ is a factor that depends on the interior structure (Methods section \ref{app:interior}). Crucially, this relationship correlates the energy flux, $q=\sigma\,T_{\rm{eff}}^4$, with the magnetic field strength, enabling the calculation of the latter from interior structure models. Arguably, the greatest source of uncertainty in a scaling law of this type corresponds to the ratio of the mean interior field to the surface field, $\gamma = \langle B \rangle/B_{\rm{s}}$. For the specific case of the Jovian dynamo, however \citet{2006GeoJI.166...97C} have shown that the ratio of interior-to-surface field strength is $\gamma \approx 6.5$, and we adopt this value as a fiducial estimate here. Staggeringly, within the range of radius derived above, the estimated surface field exhibits only a weak dependence on the planetary radius, and evaluates to $B_{\jupiter}^{\dagger}\approx21\,$mT, given our nominal estimates (Figure \ref{f3}).

It is further worth noting that the derived value of $B_{\jupiter}^{\dagger}$ should be interpreted as an effective lower bound. That is, the adopted values of $f_{\rm{ohm}}$ and $\gamma$ are close to the lower limits exhibited by dynamo models as a whole \citep{Christensen2009}, though their values tend to not vary by more than a factor of $\sim2$. Thus, while the estimate quoted in equation (\ref{yesyes}) constitutes an optimal approximation given the available data, a field strength that is twice as high (but not twice as low) is in principle plausible for the primordial Jupiter.

Assuming a dipolar field (which yields a dimensionless constant of proportionality between magnetic and accretionary pressure of $\alpha=1.96$; \citep{2008ApJ...687.1323M}), the expression for magnetospheric disk truncation takes the form:
\begin{align}
\frac{c\,f_{\rm{ohm}}\, \langle \rho \rangle^{1/3}\,\big(\mathcal{F}\,q \big)^{2/3}}{\gamma^2}\bigg( \frac{\Rjup^{\dagger}}{\Rt} \bigg)^6 = \frac{\alpha\,\dot{M}}{4\,\pi\,\Rt^2}\sqrt{\frac{2\,\G\,\Mjup}{\Rt}}.
\label{yesyesyes}
\end{align}
Applying our derived range for Jupiter's radius, we find a corresponding accretion rate of $\dot{M}\approx 1.2-2.4\Mjup$/Myr. The resulting profiles of $\dot{M}$ as well as $\Omega_{\jupiter}^{\dagger}$ and $\chi$ are shown as functions of the planetary radius in Figure \ref{f3}. We remark that in conventional disk evolution theory, the inward accretion flow in the inner disk is balanced by a much smaller outward decretion flow in the outer disk, beyond the centrifugal radius. Previous work on satellite formation \citep{BM20} indicates a value of $0.1\Mjup$/Myr for this outer decretion flow -- roughly an order of magnitude smaller than our derived estimate for the inward accretion rate.

As a check on the self-consistency and marginal improvement of our results, we recomputed the primordial Jovian radius, this time starting from the above expression to obtain $\dot{M}$ as a function of $R$ from structural models alone. Subsequently, we derived $\Rjup^{\dagger}$ by simultaneously solving for the rotational equilibrium at the end of the nebular phase (fully accounting for all terms in the angular momentum evolution equation; Methods section \ref{app:rotation}) and requiring that the terminal angular momentum matches the present-day value. This approach yields $\Rjup^{\dagger} = 2.0-2.56\,\Rjup$ as the lower and upper bounds for Jupiter's primordial radius. We adopt this refined range for the abscissa in Figure \ref{f3}, and note that the values differ from the analytical estimate (equation \ref{yes}) only by about one percent.  

\section{Methods}

\subsection{Excitation of the Inclinations of Amalthea and Thebe by Io}\label{app:dynamics}

The analysis underpinning much of this study relies upon constraints on Io's primordial semi-major axis, as established by \citet{Hamilton} through detailed N-body simulations. While a comprehensive replication of these simulations falls outside the scope of this work, it is nonetheless possible to achieve a simplified understanding of the underlying dynamics through analytic means. To this end, we will employ an integrable Hamiltonian that is based upon a literal expansion of the disturbing potential in orbital elements \citep{1999ssd..book.....M}, assuming that Io evolves on a circular, planar orbit, and neglecting any coupling between eccentricities and inclinations of the inner satellites. While this model yields results that only match the observations approximately, it does present an intelligible view of the resonant machinery behind inclination excitation of Amalthea and Thebe.

The specific calculation we undertake is centered around encounters with second-order resonances of the form $(p+2):2$, where $p$ is an integer greater than zero. Generally, such resonances feature an array of multiplets that, ideally, should to be considered collectively. However, given our assumption of massless inner satellites and initially circular and planar orbits (as well as considerations of resonance splitting induced by planetary oblateness \citep{2002mcma.book.....M}), our analysis narrows to a single resonant multiplet, characterized by the critical angle $\phi=((p+2)\,n_{\rm{Io}}\,t-p\,\lambda'-2\,\Omega')/2$, where $n_{\rm{Io}}$ is the orbital frequency of Io at the time of the resonant encounter, while $\lambda'$ and $\Omega'$ refer to the inner satellites' mean longitude and longitude of ascending node.

A minimalistic model of this resonance can be constructed by taking only the Keplerian and resonant terms of the Hamiltonian, i.e., $\mathcal{H}=\mathcal{H}_{\rm{K}}+\mathcal{H}'+\mathcal{T}$ (where $\mathcal{T}$ is a dummy action conjugate to time). To leading order, secular terms emanating from oblateness and satellite-satellite interactions only act to slightly modify the nominal semi-major axes of the resonances, and we neglect them for simplicity. Written in terms of Poincar\'e action-angle variables, the components of the Hamiltonian read:
\begin{align}
\mathcal{H}_{\rm{K}} &= -\frac{1}{2}\bigg(\frac{\G\,\M}{\Lambda'} \bigg)^2 \approx -\frac{1}{2}\bigg(\frac{\G\,\M}{[\Lambda]'} \bigg)^2 + \bigg(\frac{\G\,\M}{[\Lambda]'} \bigg)^2\,\bigg( \frac{[\Lambda]-[\Lambda]'}{[\Lambda]'} \bigg) \nonumber \\
&- \frac{3}{2} \bigg(\frac{\G\,\M}{[\Lambda]'} \bigg)^2\,\bigg( \frac{[\Lambda]-[\Lambda]'}{[\Lambda]'} \bigg)^2 \nonumber \\
\mathcal{H}'&=-\frac{\G\,\M}{2\,a_{\rm{Io}}}\,\bigg(\frac{p}{p+2} \bigg)^{2/3}\,\bigg(\frac{2\,\mathcal{Z}'}{[\Lambda]'}\bigg)\,b_{3/2}^{(p+1)}\,\cos\big((p+2)\,n_{\rm{Io}}\,t-p\,\lambda'+2\,z'\big)
\label{HHres}
\end{align}
where $b_{3/2}^{(p+1)}$ is a Laplace coefficient, $\Lambda'=\sqrt{\G\,\M\,a'}$ is the nominal specific angular momentum, $[\Lambda]'$ is the value of $\Lambda$ evaluated at exact mean-motion commensurability, $\mathcal{Z}'\approx[\Lambda]'\,(i')^2/2$ is the action related to the inner satellite's inclination, $i'$, and $z'=-\Omega'$ is the conjugated angle.

\begin{figure*}[h!]
\centering
\includegraphics[width=0.8\textwidth]{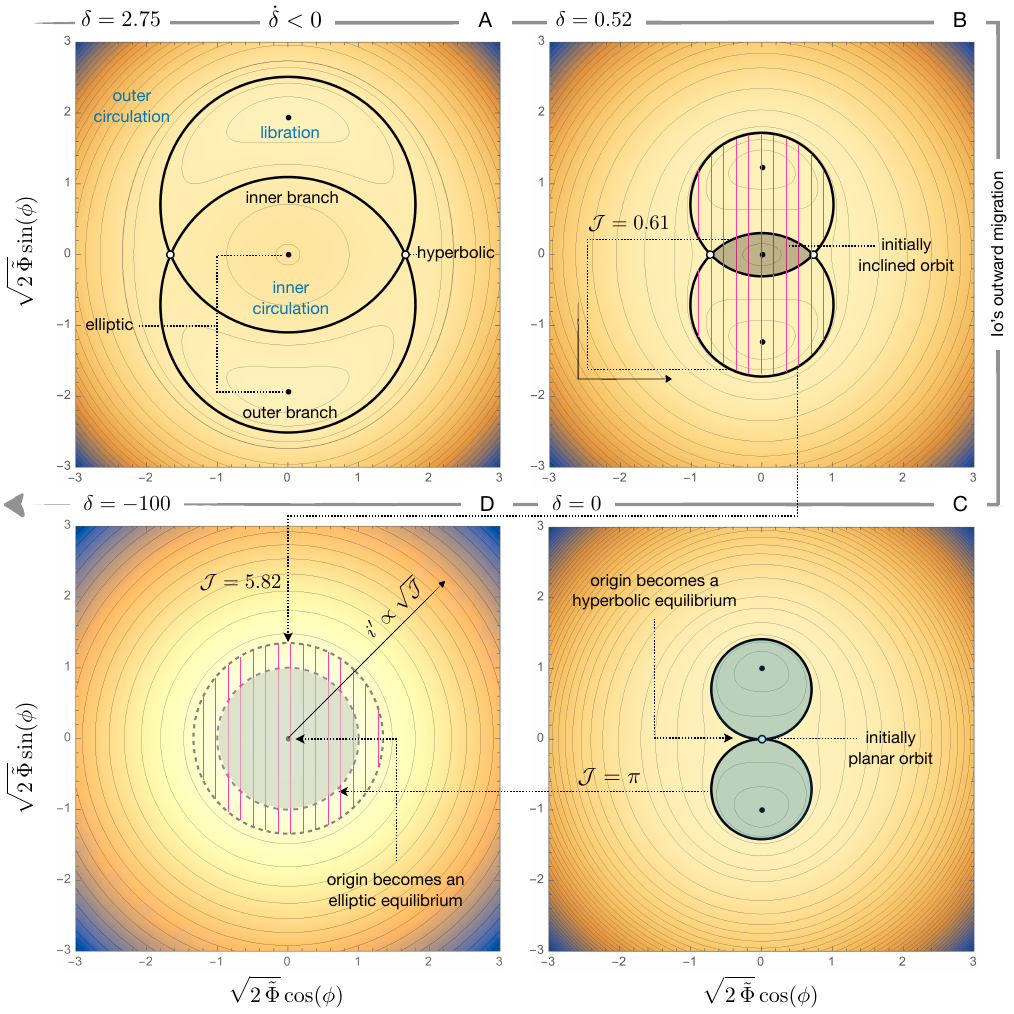}
\caption{Phase-space evolution during resonance encounters. Panels~A--D show snapshots of the Hamiltonian's phase-space portrait at different values of the proximity parameter $\delta$, which decreases as Io migrates outward. The black curves mark the separatrix, with stable (elliptic) equilibria shown as filled circles and unstable (hyperbolic) equilibria as open circles. An initially planar orbit starts at the origin ($\tilde{\Phi}=0$) and, under adiabatic evolution, conserves the phase-space area $\mathcal{J}$ (Panels~A,B). At $\delta=0$ (Panel~C), the planar equilibrium becomes hyperbolic, causing $\mathcal{J}$ to jump to the area enclosed by the separatrix. Further outward migration (Panel~D) restores adiabatic invariance with a larger $\mathcal{J}$, leading to a final inclined orbit. A single resonance crossing (with $p=1$) can account for Amalthea’s moderate inclination, whereas Thebe’s greater tilt necessitates multiple crossings (e.g., $p=4,3,2$). After the first encounter, successive encounters occur when the phase-space area of the inner branch of the separatrix matches the phase-space area occupied by the orbit, and thus occur with $\delta$ greater than 0. Colored shading in Panels~B and~C indicates successive enlargements of $\mathcal{J}$.}
\label{sfmr}
\end{figure*}

As is conventional in problems of this type, we begin with a canonical transformation of variables, through a type-2 generating function of the form $\mathcal{F}_2 = ((p+2)\,n_{\rm{Io}}\,t-p\,\lambda'+2\,z')\,\Phi/2+(\lambda')\,\Psi+(t)\,\Xi$. The transformation equations \citep{1950clme.book.....G} yield two integrals of motion -- $\Psi=\Lambda'+p\,\mathcal{Z}/2$ and $\Xi=\mathcal{T}-(p+2)\,n_{\rm{Io}}\,\mathcal{Z}/2$ -- leaving the action $\Phi=\mathcal{Z}$ and its conjugate angle $\phi$ as the sole dynamic pair of variables within the Hamiltonian. As the next step, we scale the action $\Phi$ as well as the Hamiltonian itself by a common factor
\begin{align}
\eta=\frac{16}{3}\frac{m_{\rm{Io}}}{\M}\frac{b_{3/2}^{(p+1)}}{p^{1/3}}\frac{\sqrt{\G\,\M\,a_{\rm{Io}}}}{(2+p)^{5/3}},
\label{etascale}
\end{align}
such that the coefficient upfront the square of the action term in $\mathcal{H}$ is twice as large as that of the harmonic term. Finally, we drop constant terms and scale the Hamiltonian once again by $b_{3/2}^{(p+1)}\,n_{\rm{Io}}\,(m_{\rm{Io}}/\M)\,(p/(p+2))^{1/3}$, thereby introducing slow time, and casting $\mathcal{H}$ into the form of an Andoyer Hamiltonian \citep{1984CeMec..32..127B, 2007ASSL..345.....F}:
\begin{align}
\tilde{\mathcal{H}}=(1+2\,\delta)\,\tilde{\Phi}-2\,\tilde{\Phi}^2-\tilde{\Phi}\,\cos{(2\,\phi)}.
\label{Andoyer}
\end{align}
Though the exact functional form of the proximity parameter $\delta$ is not important for the problem at hand, it is crucial to note that its dependence on Io's semi-major axis $\delta\propto (1- \sqrt{\G\,\M\,a_{\rm{Io}}}\,(p/(p+2))^{1/3}/\Psi)$ is such that $\delta$ diminishes monotonically as Io tidally recedes away from Jupiter. It is worth noting that Hamiltonians of this class serve as paradigmatic models for resonances across a broad variety of physical systems \citep{1983CeMec..30..197H}. As one example, the particle-core interaction model of halo formation in high-intensity proton beams is modeled with a Hamiltonian that is identical to equation (\ref{Andoyer}), though the physical meanings of the variables are obviously different \citep{2021PhPl...28i3104B}.

A series of phase-space portraits of the Hamiltonian -- spanning a range of values of $\delta$ -- are shown in Figure (\ref{sfmr}). In panels of this diagram that encompass hyperbolic fixed point(s), the separatrix is shown as a bold curve. In a limit where the dynamical evolution unfolds adiabatically, the phase-space area occupied by the trajectory is conserved to an excellent approximation, yielding the quasi-invariant:
\begin{align}
\mathcal{J}= \oint \tilde{\Phi}\,d\phi \approx \rm{const}.
\label{AI}
\end{align}
Even in a perfectly adiabatic system, however, the conservation of $\mathcal{J}$ can be momentarily broken if the trajectory encounters a shrinking separatrix. Qualitatively, this occurs because the separatrix is an orbit with an infinite period, rendering \textit{any} dynamics non-adiabatic with respect to its crossing. In such an event, the occupied phase-space area impulsively widens to match that of the outer branch of the separatrix, before the conservation of $\mathcal{J}$ is established again (see Figure \ref{sfmr}). Qualitatively speaking, either a single or repeated application of this process underlies resonant excitation the inner moons' inclinations. And while the eccentricities of the satellites are subject to equivalent excitations, unlike the inclinations, the eccentricities tidally decay on a timescale that is much shorter than the lifetime of the solar system.

Let us consider the case of Amalthea first, as it is somewhat simpler. Recalling that $\tilde{\Phi}\propto i'^2$, the initial value of the associated adiabatic invariant is $\mathcal{J}=0$, and remains so until the phase-space area occupied by the inner circulation region of the separatrix becomes null. This happens when the proximity parameter reaches $\delta=0$, and the corresponding phase-space area of the outer branch of the separatix is equal to $\pi$. Working backwards through the variable transformations, we arrive at the relation: $\pi\,\eta = 2\,\pi\,[\Lambda]' (i')^2/2$, which is easily rearranged to give: $i' = 4\,\sqrt{b_{3/2}^{(p+1)}\,m_{\rm{Io}}/(3\,\M\,p^{2/3}(p+2)^{4/3})}$.

For the 3:1 ($p=1$) resonant encounter between Io and Amalthea (which would have occurred at an epoch when $a_{\rm{Io}}\approx 5.27\Rjup$), our analytic theory predicts $i'=0.51\deg$ -- a value that is larger than Amaltea's true inclination of $0.37\deg$. Numerical simulations of \citet{Hamilton} that account for realistic tidal regression and non-zero inclination of Io, however, refine this estimate downward by $\sim40\,\%$, yielding a near-exact match to the observations. A similar degree of excitation is expected from an encounter with a $4:2$ resonance, which corresponds to $a_{\rm{Io}}\approx 4.02\Rjup$. Because a two-fold (i.e., 4:2 followed by 3:1 resonance) excitation of Amaltea's inclination would have resulted in $i'$ that is well above the observed value, the former could not have happened, implying that $a_{\rm{Io}}^{\dagger} > 4.02\Rjup$.

The case of Thebe is somewhat more complex, as an explanation for its $1.09\deg$ inclination is best attributed to multiple resonant encounters \citep{Burns}. To this end, let us consider sequential encounters with 6:4, 5:3, and 4:2 resonances. Assuming an initial inclination of zero, the first ($p=4$) encounter yields $i'=0.19\deg$. Setting $p=3$ for the subsequent resonance, we find that the pre-encounter value of the adiabatic invariant evaluates to $\mathcal{J}=0.61$, such that separatrix crossing ensues at $\delta=0.52$. This raises Thebe's inclination to $0.6\deg$. The final ($p=2$) encounter raises the value of $\mathcal{J}$ from $5.25$ to $15.18$, predicting a present-day inclination of $1.01\deg$ for Thebe. Though this analytic estimate is $\sim7\%$ smaller than the observed value numerical simulations \citep{Hamilton} yield closer agreement. In any case, the fact that Thebe must have crossed the 4:2 commensurability with Io restricts Io's primordial semi-major axis to $a_{\rm{Io}}^{\dagger} < 4.98\Rjup$. It is noteworthy, however, that \textit{if} a convincing case could be made for Thebe having crossed \textit{all three} resonances, then its primordial semi-major axis would be constrained to $a_{\rm{Io}}^{\dagger} < 4.06\Rjup$, in turn yielding an exceptionally narrow range for Jupiter's primordial radius.

\subsection{Disk-Driven Orbital Migration Near the Magnetospheric Cavity}\label{app:migration}

An inescapable consequence of satellite accretion within a circumplanetary disk of gas and dust is the manifestation of satellite-disk interactions \citep{1980ApJ...241..425G}. For objects with masses below the gap-opening threshold these interactions ensue in the linear (or the so-called Type-I) regime. With normalized masses in the range of $m/M=2.5 -7.8 \times10^{-5}$, the Galilean satellites fall squarely into this category \citep{2006Icar..181..587C}, while simultaneously being massive enough to not succumb to stochastic evolution facilitated by disk turbulence \citep{2008ApJ...683.1117A,2017AJ....153..120B,2024MNRAS.528L.127W}. The associated rate of orbital migration can be expressed as:
\begin{align}
\frac{1}{\tau_{\rm{mig}}} = -\frac{\dot{r}}{r} = \hat{\Upsilon}\,\bigg( \frac{m}{M} \bigg) \,\bigg(\frac{\Sigma \, r^2}{M} \bigg) \,\bigg(\frac{r}{h} \bigg)^2\,\sqrt{\frac{\G\,\Mjup}{r^3}}
\label{taumig}
\end{align}
where $\hat{\Upsilon}$ is a factor that depends on the structure of the disk. More specifically, $\hat{\Upsilon} = \sum \Upsilon$ represents the sum over contributions from a series of effects (e.g., Linblad torque, horseshoe drag., etc.) that collectively drive orbital evolution, such that if $\hat{\Upsilon}$ is positive, the sense of migration is inward. Typically, the value of $\hat{\Upsilon}$ falls between 0 and 1, meaning that inward migration ensues at a rate that is somewhat slower than that implied solely by the baseline dependencies in equation (\ref{taumig}). Given the decretion disk model of \citet{BM20}, the fiducial orbital decay time of Io evaluates to $\hat{\Upsilon} \, \tau_{\rm{mig}} \sim 10^4\,$years at $r\sim10^2 \Rjup\gg\Rt$.

In the vicinity of the truncation radius, the situation chances considerably, as the direction of type-I migration direction reverses (such that the satellite is repelled outward from the disk's edge; \citep{Masset2006}). Mathematically, this occurs due to fact that barotropic contributions to $\hat{\Upsilon}$ from the horseshoe drag and the linear corotation torque are sensitive to the local power-law index, $\beta$, of the surface density profile (i.e., $\Sigma\, \propto\, r^{-\beta}$), and scale as:
\begin{align}
\Upsilon_{\rm{baro}}\, \propto\, \bigg( \frac{3}{2}-\beta \bigg).
\end{align}
Because $\beta\ll0$ at the disk's inner edge, the outward-directed torque is greatly amplified, dominating over other contributions and yielding $\hat{\Upsilon}<0$. From a more intuitive point of view, this torque reversal can be understood by envisioning satellite-driven co-orbital dynamics of the gas. Particularly, at the disk's periphery, nebular fluid is shuttled from denser outer regions to less dense inner areas along the horseshoe orbit. Owing to conservation of angular momentum, this form of satellite-gas interactions must facilitate outward migration.

Ultimately, the equilibrium semi-major axis -- where competing torques nullify each other -- must exceed the truncation radius by a small amount. Using a series of detailed calculations, \citet{AtaieeKley21} have  demonstrated that, for resonant chains comprising three objects with mass ratios of $m/M=3\times10^{-5}$, $3\times10^{-5}$, and $6\times10^{-5}$ (similar to the Io-Europa-Ganymede triplet), this equilibrium radius surpasses $\Rt$ by a factor of $\zeta \approx 1.13$. Crucially, their analysis reveals a robust outcome: irrespective of substantial (i.e., order-of-magnitude) variations in masses, all resonant chains, whether comprised of two or three bodies, stabilize with $\zeta$ ranging between 1.1 and 1.2. These results are further consistent with the hydrodynamical simulations of \citet{Masset2006}, who also find $\zeta \approx 1.13$ for $m/M\approx0.4-5.9\times10^{-5}$. 

\subsection{Rotational Angular Momentum}\label{app:rotation}

During the disk-bearing phase, the spin angular momentum of the planet is modulated by magnetohydrodynamic coupling between the planetary magnetic field and the disk plasma. In the high magnetic Reynolds number regime (applicable to the inner $\sim$third of the circum-Jovian disk; \citep{Bat18}), the Lorentz torque associated with a pole-aligned dipole field reads \citep{1992MNRAS.259P..23L}:
\begin{align}
\Lambda = - \frac{4\,\pi}{\mu_0}\int \varphi\, \frac{\mathcal{M}^2}{r^4}\, dr &= \begin{dcases} -\frac{4\,\pi}{3\,\mu_0}\frac{\mathcal{M}^2}{R_t^3} & \omega > \sqrt{\frac{\G\,M}{R_t^3}} \\ \frac{4\,\pi}{3\,\mu_0}\frac{\mathcal{M}^2 \, \big(R_c^3-2\,R_t^3 \big)}{R_t^3\,R_c^3} & \omega \leqslant \sqrt{\frac{\G\,M}{R_t^3}},\end{dcases}
\label{tautrue}
\end{align}
where $\mathcal{M}=B_{\jupiter}^{\dagger}\,\Rjup^3$ is the magnetic moment, and $\varphi = B_{\theta}/B_{z}$ is the pitch angle of the field within the disk. Strictly speaking, the integral in equation (\ref{tautrue}) runs over the disk's magnetically connected region i.e., from $\Rt$ to approximately one fifth of the Jovian Hill sphere \citep{Bat18}. Given that the integrand diminishes quickly with planetocentric distance, however, the outer boundary can be taken as being effectively infinite to an excellent approximation. In terms of magnitude, magnetic reconnection maintains $|\varphi |\approx 1$, but its sign reverses across the corotation radius, $\Rc=(G\,\Mjup/\Omega^2)^{1/3}$, such that induction leads to stellar spin-down outside of $\Rc$, and spin-up interior to $\Rc$ \citep{2002ApJ...565.1205U}. Technically, there exists a radially thin zone, adjacent to the corotation radius, where Keplerian shear is weak enough for Ohmic diffusion to sustain $\varphi<1$. Despite its presence, this sector is minuscule and can be disregarded with minimal consequence to our analysis \citep{2005MNRAS.356..167M}.

It is easy to see that $\Lambda$ vanishes when the corotation radius matches an equilibrium value of $\Rc=2^{1/3}\, \Rt$ (this configuration is sometimes referred to as the disk-locked state). Acting in isolation, the Lorentz torque would drive the planetary spin toward this equilibrium over a characteristic time-interval:
\begin{align}
\tau_m = \frac{3\,\mu_0\,I_{\jupiter}^{\dagger}\,\Rt^3\,\sqrt{\G\,\Mjup^3\,\Rjup}}{4\,\pi\,\mathcal{M}^2}\lesssim10^5\,\rm{yr}.
\label{taum}
\end{align}
Indeed, this is the shortest timescale associated with rotational evolution of giant planets \citep{Bat18}.

While dominant, the Lorentz torque is not the only effect that influences a planet's rotational evolution. Gravitational contraction and accretion can facilitate significant spin-up, nudging $\Omega$ away from the disk-locked state described above. The full 1D evolutionary equation is thus given by \citep{1996MNRAS.280..458A}: 
\begin{align}
\dot{J}&=\dot{I}_{\jupiter}^{\dagger}\,\Mjup\,\Rjup^2\,\Omega_{\jupiter}^{\dagger}+I_{\jupiter}^{\dagger}\,\dot{M}\,\Rjup^2\,\Omega_{\jupiter}^{\dagger} + 2 I_{\jupiter}^{\dagger}\,\Mjup\,\Rjup\,\dot{R}_{\jupiter}^{\dagger}\,\Omega_{\jupiter}^{\dagger} \nonumber \\
&+ I_{\jupiter}^{\dagger}\,\Mjup\,\Rjup^2\,\dot{\Omega}_{\jupiter}^{\dagger} = \Lambda + \dot{M}\,\sqrt{\G\,\Mjup\,\Rt}.
\label{Jdotfull}
\end{align}
The solution to equation (\ref{Jdotfull}) follows a well-defined characteristic: over a timescale akin to $\tau_m$, the system tends towards a state where the accelerative spin-up due to accretion and gravitational contraction finds its counterweight in the decelerative effect of magnetic braking, yielding $\dot{\Omega}_{\jupiter}^{\dagger}\rightarrow0$. Interestingly, the terms on the left-hand side (LHS) of this equation provide only minor (order percent-level, as can be demonstrated from a simple timescale analysis based upon quantities delineated in Table \ref{tab1}), corrections to the balance of terms on the RHS. In other words, the equilibrium planetary spin is, to a good approximation, determined by a competition between accretion and Lorentz torques alone.

Fortuitously, the equilibrium spin derived from from accretion and magnetic braking is independent of the planetary radius or the magnetic field's strength. This stems from the fact that in the well-known disk truncation formula \citep{GhoshLamb78}:
\begin{align}
\frac{\mathcal{M}^2}{2\,\mu_0\,\Rt^6} = \frac{\alpha\,\dot{M}}{4\,\pi\,\Rt^2}\sqrt{\frac{2\,\G\,\Mjup}{\Rt}},
\label{truncationradiuseqn}
\end{align}
$\dot{M}$ scales with the square of the magnetic moment — a dependency mirrored in the Lorentz torque equation (\ref{tautrue}) (this cancellation is merely a reflection of the common origin between magnetic disk-truncation and magnetic breaking). Accordingly solving for the equilibrium rotational frequency, we obtain:
\begin{align}
\Omega_{\jupiter}^{\dagger}=\Bigg(\frac{1}{2}\,\sqrt{2+\frac{3}{\sqrt{2}\,\alpha}} \Bigg)\,\sqrt{\frac{\G\,\Mjup}{\Rjup^3}} = \chi\,\Omega_t.
\label{Omegajup}
\end{align}
Adopting the value of the dimensionless constant $\alpha=1.96$ derived by \citet{2008ApJ...687.1323M} for a pole-aligned dipolar field, the factor in the parenthesis evaluates to $\chi=0.88$. This implies that accounting for accretion, Jovian spin frequency exceeds the disk-locked equilibrium by a factor of 1.24.

\subsection{Interior Structure Models}\label{app:interior}

A critical step in determining Jupiter's primordial radius from the conservation of angular momentum (equation \ref{yes}) involves the calculation of its moment of inertia factor $I=J/(M\,R^2\,\Omega)$. Thanks to data from the Juno mission, the present-day moment of inertia of Jupiter, $I_{\jupiter}$ is rather well-established, with derived values ranging from $0.264$ \citep{Helled2011MOI,Wahl2017} to $0.276$ \citep{Ni2018}. Shortly after formation, however, Jupiter had a more centrally concentrated density profile, resulting in a substantially lower moment of inertia factor \citep{Helled2012}.

To compute $I_{\jupiter}^{\dagger}$ within the radius range pertinent to the epoch of disk dissipation, we employed the \texttt{MESA} software package \citep{2013ApJS..208....4P} to generate a series of hydrostatic models with $M=\Mjup$ and radii ranging from $1.5\,\Rjup$ to $3\,\Rjup$. These models assume a solar-composition gaseous envelope and incorporate a heavy-element core with a density of $\rho_{\rm{c}}=10\,$g/cc and a mass of $M_{\rm{c}} = 25 M_{\oplus}$, aligning with the total inferred heavy-element content of the planet \citep{MH24}. Our runs employed the default \texttt{MESA} equation of state (EOS), which is a blend of several tables appropriate for different regions of parameter space. For the planetary regime, this equation of state is approximately equivalent to the \texttt{SCVH} \citep{1995ApJS...99..713S} model. The atmosphere model is based on the grey approximation, and opacities were taken from the \texttt{gs98} \citep{1998SSRv...85..161G} tables for high temperatures and \texttt{lowT\_Freedman11} tables \citep{2008ApJS..174..504F} for low temperatures, with a base metallicity of $0.02$.

We note that recent advances in EOS of hydrogen-helium mixtures \citep{2019ApJ...872...51C} provide significant improvements in the high-density (i.e., $\rho\gtrsim 1\,$g/cc) regime. These advancements have already proved important for the present-day Jupiter -- where central densities are high and the choice of EOS significantly influences modeling outcomes \citep{2004ApJ...609.1170S, 2016A&A...596A.114M, 2022A&A...664A.112M, 2023A&A...672L...1H}. In the early stages of Jupiter's evolution, interior densities within the envelope are considerably lower, and temperatures are significantly higher, indicating that the sensitivity to the specific method of EOS calculation is somewhat reduced. Nevertheless, future work should improve this, and other aspects of interior modeling, including the introduction of a dilute core (see e.g., refs \cite{2023A&A...672A..33H,2024arXiv240709341K}), to further enhance the accuracy of the calculations.

\begin{figure*}[t]
\centering
\includegraphics[width=0.80\textwidth]{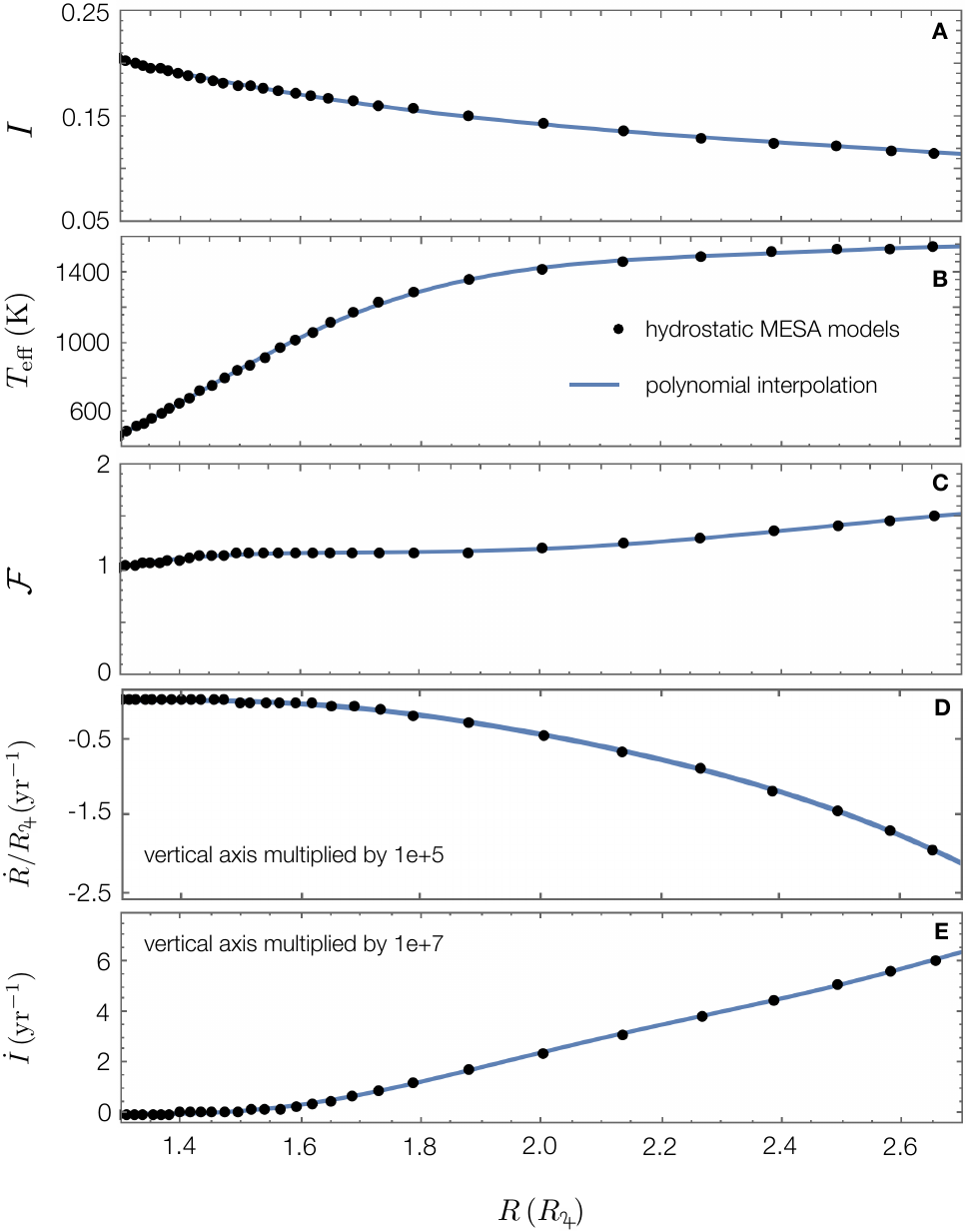}
\caption{Quantities derived from hydrostatic interior models, shown as functions of Jupiter's radius. Each panel shows black points that correspond to \texttt{MESA} output, with the blue line denoting a polynomial interpolation. Panel A depicts the dimensionless moment of inertia factor $I$. Panel B shows the atmospheric effective temperature $(T_{\mathrm{eff}})$. Panel C illustrates the structural coefficient $\mathcal{F}$, which appears in the dynamo scaling law. Panel D gives the fractional contraction rate $\dot{R}_{\jupiter}/R_{\jupiter}$ (with vertical axis multiplied by $10^{5}$). Panel E displays the rate of change of the dimensionless moment of inertia, $\dot{I}$ (with vertical axis multiplied by \(10^{7}\)).}
\label{fmesa}
\end{figure*}

\begin{table}[t]
\caption{A sequence of interior structure models.}\label{tab1}%
\begin{tabular}{@{}llllll@{}}
\toprule
$R$ ($\Rjup$) & $I_{\jupiter}^{\dagger}$  & $T_{\rm{eff}}$ (K) & $\mathcal{F}$ & $\dot{R}/\Rjup$ (yr$^-1$) & $\dot{I}$ \\
\midrule
2.49    & 0.122   & 1528  & 1.43  & -1.4\,e-5 & 5.1\,e-7 \\
2.39    & 0.126   & 1515  & 1.38 & -1.1\,e-5 & 4.5\,e-7\\
2.26 	  & 0.131 & 1496 & 1.31  & -8.7\,e-6 & 3.9\,e-7 \\
2.13 & 0.137 & 1467 & 1.25 & -6.4\,e-6 & 3.2\,e-7 \\
2.00 & 0.143 & 1428 & 1.21 & -4.4\,e-6 & 2.4\,e-7 \\
1.94    & 0.145$^{\ast}$   & 1440  & -- & -- & --   \\ 
\botrule
\end{tabular}
\footnotetext[\ast]{Derived from the digitized density profile of an interior model from \citet{Stevenson22}.}
\end{table}

We calculate the moment of inertia directly from the density profiles as follows:
\begin{align}
I_{\jupiter}^{\dagger} = \frac{2}{3}\frac{1}{M\,R^2}\int_{0}^{R}4\,\pi\,\rho\,r^4\,dr.
\label{MOI}
\end{align}
 The resultant values of $I_{\jupiter}^{\dagger} $, alongside the effective temperatures ($T_{\rm{eff}}$) and other quantities are detailed in Table \ref{tab1} for a representative series of interior models, and Figure \ref{fmesa} depicts them a function of the planetary radius. As a means of verifying that our derived value of $I_{\jupiter}^{\dagger}$ is not sensitive to the assumption of a solid core, we digitized the density profile of the diffuse core $R\approx2\Rjup$ model of \citet{Stevenson22} and computed the moment of inertia from their data. This check gave consistent results: while the \citet{Stevenson22} model gives $I_{\jupiter}^{\dagger} = 0.145$, the \texttt{MESA} models adopted herein give $I_{\jupiter}^{\dagger} = 0.147$ for the same radius. 
 
 Additionally, for each model, we calculated the structural coefficient, integral to the dynamo scaling law \citep{Christensen2009}:
\begin{align}
\mathcal{F}^{2/3} = \frac{3}{4\,\pi\,R_{\rm{d}}^3} \int_{R_{\rm{c}}}^{R_{\rm{d}}} \bigg(\frac{\mathcal{L}}{\mathcal{H}_{T}} \frac{q}{q_{\rm{surf}}} \bigg)^{2/3} \bigg( \frac{\rho}{\langle \rho \rangle}  \bigg)^{1/3} \,4\,\pi\,r^2\,dr.
\label{Feqn}
\end{align}
In the above expression, $\mathcal{H}_{T}$ is the temperature scale-height, $\mathcal{L} = \min(\mathcal{H}_{\rho}, R-r)$ is the inferior of the density scale-height and the distance from $r$ to the edge of the convective region (for the relevant models the radiative-convective boundary is close to the photosphere), $q$ is the flux, and $q_{\rm{surf}}=\sigma\,T_{\rm{eff}}^4$. Notably, the integral runs from the core radius to the top of the dynamo region, which we take to correspond to the onset of large-scale ionization of Hydrogen. For definitiveness, we take the top of the dynamo region as the radius at which the mean number of free electrons per nucleon exceeds $3\times10^{-4}$. Though somewhat arbitrary, our results are not particularly sensitive to this assumption.

For the relevant planetary radii, we found the integral (\ref{Feqn}) to lie in the range of $\mathcal{F}=1.21-1.43$ (see Table {\ref{tab1}}). The closeness of $\mathcal{F}$ to unity is not surprising: Christensen et al. \citep{Christensen2009} have emphasized that this coefficient, reflecting the efficiency of magnetic field generation, is on the order of one across a wide spectrum of astrophysical objects. In other words, this quantity is rather insensitive to the exact density profile of the interior, and it is unlikely that modeling a diffuse core would appreciably alter the value of $\mathcal{F}$.
  
\subsection{Orbital Decay Within the Magnetospheric Cavity}\label{app:stokes}

Traditional models of a disk's magnetospheric cavity (e.g., \citep{GhoshLamb78}) envision a scenario where disk plasma ascends along the critical field line at the truncation radius, forming an accretion ``curtain" that feeds the central body. As such, the region interior to this field line has routinely been approximated as having a drastically reduced surface density. Recent simulations of \citet{Zhu24}, however, have introduced a more complex picture, illustrating that while some of the accretionary flow ultimately gets carried away by winds \citep{1994ApJ...429..797S}, a fraction of the disk plasma can infiltrate the magnetospheric cavity by way of the interchange instability, cascading towards Jupiter in a free-fall-like manner. This dynamic results in a sparse, but noticeably non-zero density profile inside the truncation radius, approximately given by:
\begin{align}
\rho = \frac{1}{\sqrt{2}}\frac{\varpi\,\dot{M}}{4\,\pi\,r^2}\frac{1}{v_{\rm{K}}}\bigg(1-\frac{r}{\Rt} \bigg)^{-1/2},
\label{rho}
\end{align}
where the square-root factor accounts for the fact that infall originates at $\Rt$ rather than infinity, and $\varpi\leqslant 1$ is the fraction of the accretionary flow that perforates the cavity.

Within the truncation radius, the plasma, to a first approximation, co-rotates with Jupiter, imparting an azimuthal velocity of $v_{\theta}= \Omega_{\jupiter}^{\dagger}\,r$. Consequently, macroscopic objects residing on nearly-Keplerian trajectories, will experience orbital decay due to Stokes drag. The associated characteristic timescale, $\tau_{\rm{aero}}$, is well-known \citep{1976PThPh..56.1756A}, and upon substituting equation (\ref{rho}) for the gas density and adopting $\nu=(\Omega_{\jupiter}^{\dagger}\,r - v_{\rm{K}})/v_{\rm{K}}$ as the sub-Keplerian factor, we obtain:
\begin{align}
\tau_{\rm{aero}} = \frac{64\,\pi\,s_{\bullet}\,\rho_{\bullet}}{3\,\mathcal{C}_{\rm{D}}\,\varpi\,\dot{M}} \bigg(r\,\sqrt{2-\frac{2\,r}{\Rt}} \, \bigg(\frac{r}{\Rt}\,\sqrt{\frac{2}{\Rt}}-\sqrt{\frac{4}{r}} \bigg)^{-2} \bigg),
\label{rho}
\end{align}
where $\rho_{\bullet}=1\,$g/cc is the assumed density of the satellites, $s_{\bullet}$ is the physical radius, and $\mathcal{C}_{\rm{D}}=0.44$ is the drag coefficient, appropriate for high-Reynolds number flow.

Adopting our lower-bound accretion rate of $\dot{M}=1.2\Mjup/$Myr and $\varpi=1/2$ for definitiveness, we obtain $\tau_{\rm{aero}} \approx 0.14\,$Myr and $0.16\,$Myr for Amalthea and Thebe (evaluated on their present-day orbits) respectively. If we envision a scenario where Amalthea and Thebe were delivered into the magnetospheric cavity by way of resonant transport, it is natural to expect that both of these moons would have undergone significant orbital decay within the cavity (by comparison, the survival of Adrastea and Metis, within the magnetospheric cavity appears improbable due to their small physical radii, hinting at the possibility that they achieved their present-day orbits after nebular dissipation). The estimated timeframe further suggests that Io's arrival to the disk's inner edge would have occurred towards the end of the nebula's lifetime.

Intriguingly, this timeline aligns with current models of circumplanetary disk evolution. Notably, hydrodynamical simulations reported in refs. \citep{2016MNRAS.460.2853S,2019A&A...630A..82L, 2024arXiv240214638K} propose that circumplanetary disks emerge relatively late within the evolutionary sequence of a giant planet, i.e., once the planet cools sufficiently (see also the analytic treatment of \citet{2022ApJ...934..111A}). Given that gravitational contraction occurs rapidly (published models suggest that post-runaway, Jupiter reaches $R<2\,\Rjup$ in a matter of hundreds of thousands of years; \citep{2005Icar..179..415H,2021Icar..35514087D,Stevenson22}), the emergent picture suggests that Jupiter's phase of rapid gas accretion ensued well within the last Myr of the solar nebula's lifetime. This is further consistent with rapid ($\lesssim 100\,$kyr) moon accretion observed within the simulations of \citet{BM20}. Cumulatively, these estimates posit that the aerodynamic drift timescales of Amalthea and Thebe are consistent with published results, presenting a coherent narrative for the  evolution of Jupiter's inner moons.

\bmhead{Data availability} Ascii \texttt{MESA} output files summarizing the interior profiles shown in Figure \ref{f2} are available for download at https://www.konstantinbatygin.com/jupiter.

\bmhead{Code availability} This work utilizes the \texttt{MESA} stellar evolution code, publicly available at https://docs.mesastar.org/.

\bmhead{Acknowledgements} K. B. is grateful to Caltech, the David and Lucile Packard Foundation, and the National Science Foundation (grant number: AST 2408867) for their generous support. F.C.A. is supported in part by the University of Michigan and the Leinweber Center for Theoretical Physics. K.B. and F.C.A. thank the referees for providing thorough and insightful reports. 

\bmhead{Author contributions} K.B. conceived the project, ran the MESA simulations, and led the writing of the manuscript. F.C.A. collaborated on the interpretation of the results, provided critical feedback on the methodology, and contributed to writing and revising the manuscript.

\bmhead{Competing Interests} The authors declare no competing interests.

\end{document}